\documentclass[10pt,letterpaper]{article}
\usepackage[top=0.85in,left=2.75in,footskip=0.75in]{geometry}

\usepackage{amsmath,amssymb}
\usepackage{pdfpages}
\usepackage{changepage}

\usepackage[utf8x]{inputenc}

\usepackage{textcomp,marvosym}

\usepackage{cite}

\usepackage{nameref,hyperref}

\usepackage[right]{lineno}

\usepackage{microtype}
\DisableLigatures[f]{encoding = *, family = * }

\usepackage{array}

\newcolumntype{+}{!{\vrule width 2pt}}

\newlength\savedwidth

\newcommand\thickhline{\noalign{\global\savedwidth\arrayrulewidth\global\arrayrulewidth 2pt}%
\hline
\noalign{\global\arrayrulewidth\savedwidth}}

\raggedright
\usepackage{dirtytalk}
\usepackage{amssymb, amsmath}
\usepackage{bbm}
\usepackage{float}
\usepackage[aboveskip=1pt,labelfont=bf,labelsep=period,justification=raggedright,singlelinecheck=off]{caption}

\makeatletter
\renewcommand{\@biblabel}[1]{\quad#1.}
\makeatother

\usepackage{lastpage,fancyhdr,graphicx}
\usepackage{epstopdf}
\fancyheadoffset[L]{2.25in}
\fancyfootoffset[L]{2.25in}
\begin{document}
\vspace*{0.2in}

\begin{flushleft}
{\Large
\textbf\newline{Nowcasting with leading indicators applied to COVID-19 fatalities in Sweden} 
}
\newline
\\
Fanny Bergström\textsuperscript{1,*},
Felix G{\"u}nther\textsuperscript{1},
Michael Höhle\textsuperscript{1},
Tom Britton\textsuperscript{1}
\\
\bigskip
\textbf{1} Department of Mathematics, Stockholm University, Stockholm, Sweden
\\
\bigskip

%
%





* fanny.bergstrom@math.su.se

\end{flushleft}
\section*{Abstract}
The real-time analysis of infectious disease surveillance data, e.g., in the form of a time-series of reported cases or fatalities, is essential in obtaining situational awareness about the current dynamics of an adverse health event such as the COVID-19 pandemic. This real-time analysis is complicated by reporting delays that lead to underreporting of the number of events for the most recent time points (e.g., days or weeks). This can lead to misconceptions by the interpreter, e.g., the media or the public, as was the case with the time-series of  reported fatalities during the COVID-19 pandemic in Sweden. Nowcasting methods provide real-time estimates of the complete number of events using the incomplete time-series of currently reported events by using information about the reporting delays from the past. Here, we consider nowcasting the number of COVID-19-related fatalities in Sweden. We propose a flexible Bayesian approach, extending existing nowcasting methods by incorporating regression components to accommodate additional information provided by leading indicators such as time-series of the number of reported cases and ICU admissions. By a retrospective evaluation, we show that the inclusion of ICU admissions as a leading signal improved the nowcasting performance of case fatalities for COVID-19 in Sweden compared to existing methods.

\section*{Author summary}
Nowcasting methods are an essential tool to provide situational awareness in a pandemic. The methods aim to provide real-time estimates of the complete number of events using the incomplete time-series of currently reported events and the information about the reporting delays from the past. In this paper, we consider nowcasting the number of COVID-19 related fatalities in Sweden. We propose a flexible Bayesian approach, extending existing nowcasting methods by incorporating regression components to accommodate additional information provided by leading indicators such as time-series of the number of reported cases and ICU admissions. We use a retrospective evaluation covering the second (alpha) and third (delta) wave of COVID-19 in Sweden to assess the performance of the proposed method. We show that the inclusion of ICU admissions as a regression component improved the nowcasting performance (measured by the CRPS score) of case fatalities for COVID-19 in Sweden by 4.2\% compared to an established method.

\newpage
\section*{Introduction}
\begin{figure}[!h]
\includegraphics[width=1\textwidth]{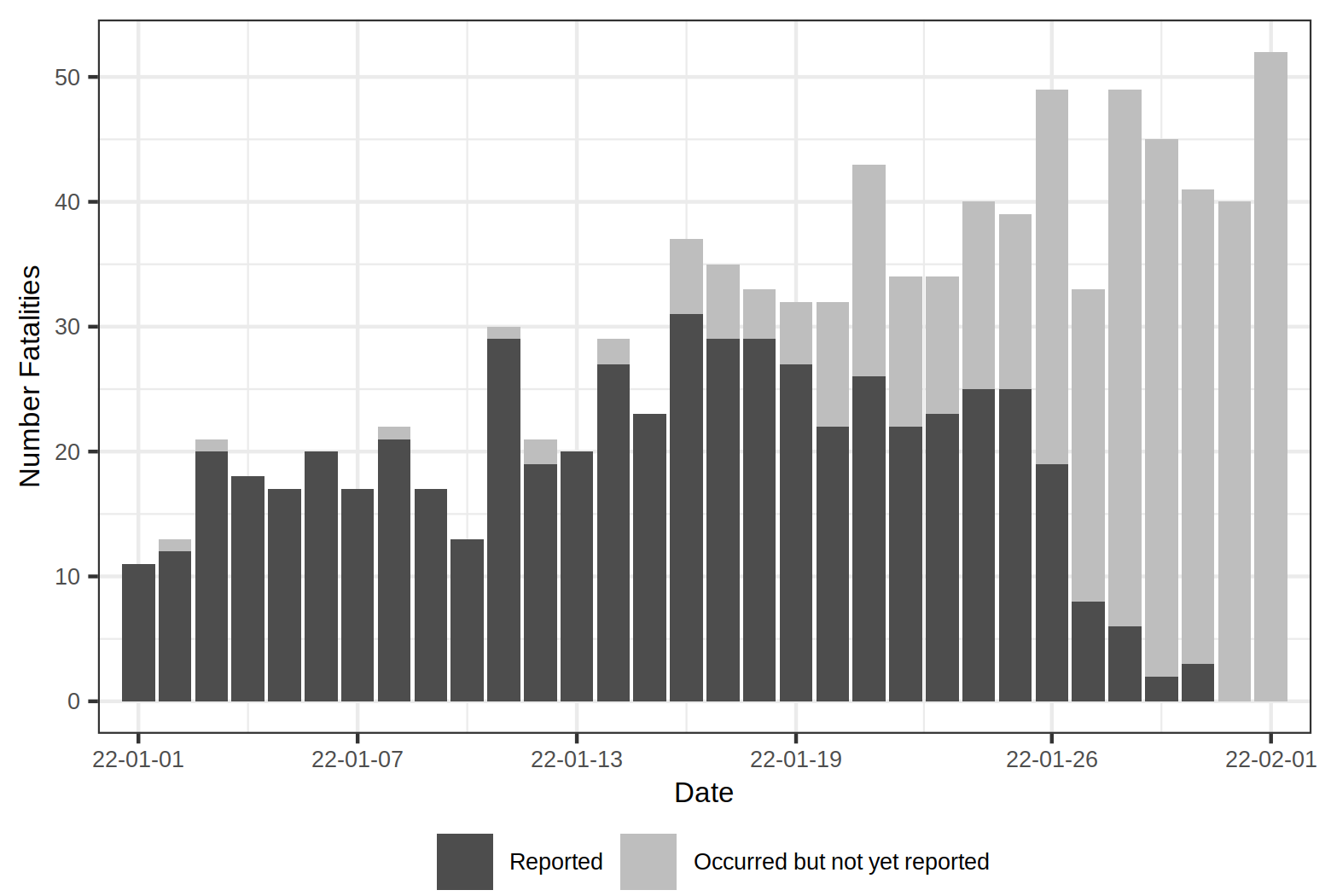}
\vspace{0.1mm}
\caption{{\bf Daily COVID-19 fatalities in Sweden.}
Reported (black bars) and unreported (grey bars) number of daily fatalities as of 2022-02-01. The reported number of events show a declining trend when in actuality (known in hindsight) it was increasing.}
\label{fig:rep_vs_unrep}
\end{figure}

The real-time analysis of infectious disease surveillance data, e.g. in the form of time-series of reported cases or fatalities, is one of the essential components in shaping the response during infectious disease outbreaks such as major food-borne outbreaks or the COVID-19 pandemic. Typically, public health agencies and governments use this type of monitoring to assess the disease dynamics and plan and assess the effectiveness of preventive actions \cite{Metcalf_2020, wu_2021}. Such real-time analysis is complicated by reporting delays that give rise to \textit{occurred-but-not-yet-reported} events which may lead to underestimation of the complete number of reported events \cite{lawless_1994}. Fig~\ref{fig:rep_vs_unrep} illustrates the problem with data of Swedish COVID-19-related fatalities as of 2022-02-01, where the reported number of fatalities per day shows a declining trend. With data available two months later \cite{FHM}, it is seen that the number of fatalities per day was at the time increasing.

Nowcasting methods \cite{donker_etal2011,hohle_2014, mcgough_etal2020} tackle this problem by providing real-time estimates of the complete number of events using the incomplete time-series of currently observed events and information about the reporting delay from the past. 
The methods have connections to insurance claims-reserving \cite{kaminsky_1987}, and its epidemiological applications trace back to HIV modelling \cite{kalbfleisch1989inference, zeger_etal1989, lawless_1994}. Nowcasting methods has been used in COVID-19 analysis for daily infections \cite{greene_2021, Tenglong_2021, seaman_2022}, and fatalities \cite{Schneble_2020, altmejd_2020, bird_2021}. A Bayesian approach to nowcasting, which constitutes the foundation of our method, was developed by Höhle and an der Heiden \cite{hohle_2014} and later extended by G{\"u}nther et al.~\cite{gunther_2020} and McGough et al. \cite{mcgough_etal2020}. Most nowcasting methods are focused on estimating the reporting delay distribution; however, an epidemic contains a temporal dependence since --depending on the mode of transmission--  adheres to certain \say{laws}, e.g. contact behavior which only changes slowly. Taking the temporal dependence of the underlying disease transmission into account has been shown to improve the nowcasting performance \cite{mcgough_etal2020, gunther_2020}. 
Another approach to nowcasting, not considering the reporting delay distribution, is to use other data sources that are sufficiently correlated with the time series of interests,  e.g.\ the Machine Learning approach by Peng et al.\cite{peng_2021}.

Our approach for nowcasting Swedish COVID-19 fatalities is based on a Bayesian hierarchical model that can account for temporal changes in the reporting delay distribution and, as an extension to existing methods~\cite{hohle_2014, gunther_2020}, incorporates a regression component of additional correlated data streams. Here, we consider the following two additional data streams; the time-series of the number of Intensive Care Unit (ICU) admissions and reported cases. The disease stages (infected, hospital, ICU, death) have a time order, and the number of new entries in one of the earlier compartments can help estimate what will happen for the later stages. As the additional data streams are assumed to be ahead in time, we consider them as leading indicators for the event of interest.

In this paper, we present methodological details of our approach and compare the results to existing nowcasting methods to illustrate the implication of incorporating additional data streams associated with the number of fatalities. We show with a retrospective evaluation of our method that nowcasting with leading indicators can improve performance compared to existing methods.

\section*{Materials and methods}
\subsection*{Data}
\begin{figure}[!h]
\includegraphics[width=1\textwidth]{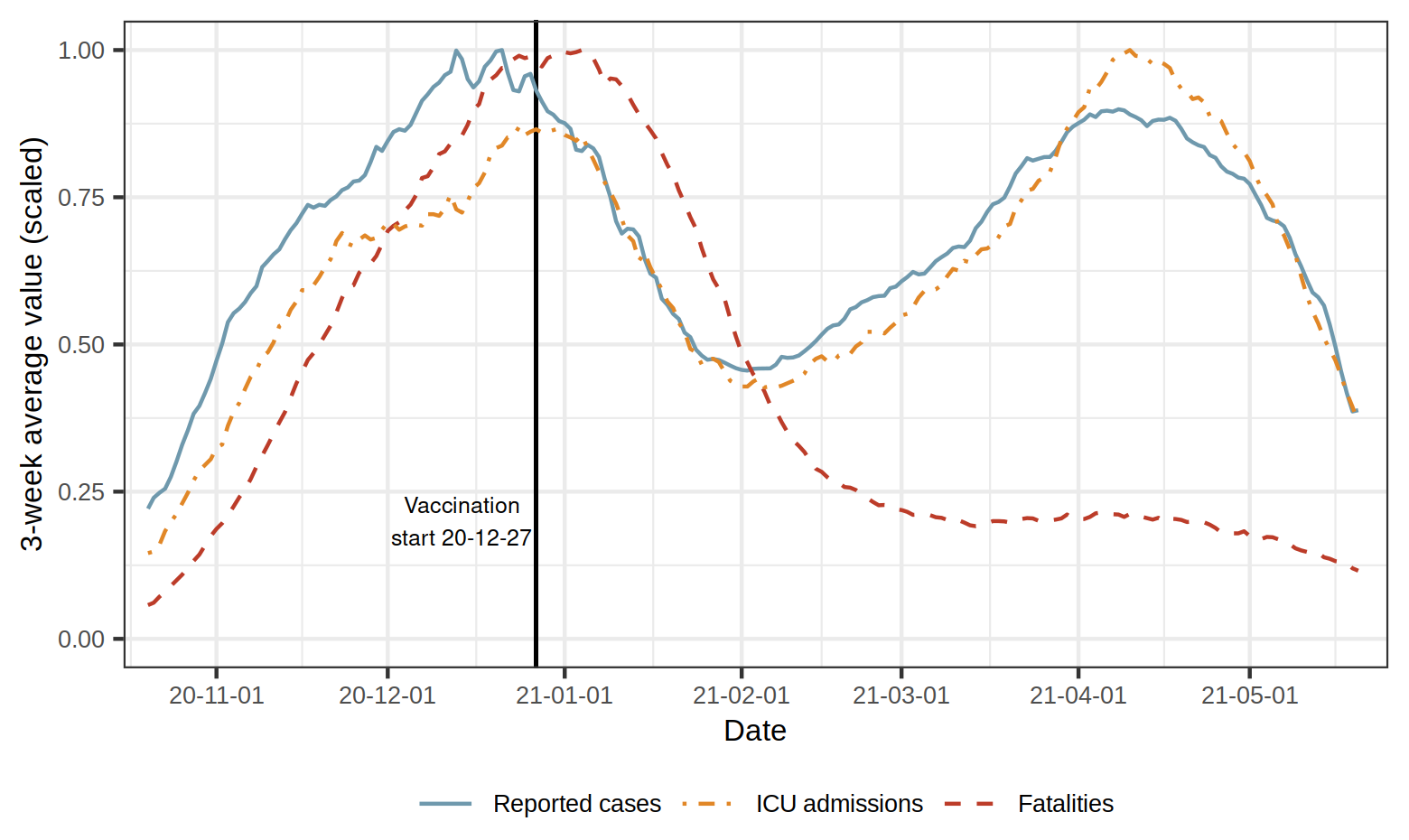}
\caption{{\bf Reported cases, ICU admissions and fatalities with COVID-19 in Sweden.}
The period covers the second (alpha) and third (delta) wave and the start of vaccination in Dec 2020. Each time series is shown with a 3-week centered rolling average and scaled by its maximum value.}
\label{fig:timeseries}
\end{figure}

The surveillance data used for the analysis in this paper are daily counts of fatalities and ICU admissions and reported cases of people with a laboratory-confirmed SARS-CoV-2 infection in Sweden. The chosen period ranges from 2020-10-20 to 2021-05-21 and contains 117 reporting days (Tuesday to Friday excluding public holidays). During this period, there were 951 646 reported cases, 4 734 ICU admissions and 8 656 fatalities. The evaluation period covers Sweden's second (alpha) and third wave (delta) of COVID-19-related fatalities. In addition, this period also covers the introduction of vaccination which meant a change in the association between reported cases or ICU admissions and the fatalities. The times series of the number of reported cases, ICU admissions, and deaths can be seen in Fig~\ref{fig:timeseries}. The figure shows that the rise and fall of the three time series follow a similar time trend, with some time delay, during the first wave. However, in the second wave, the relative association between the fatalities and the other disease stages becomes less substantial, the main reason being the introduction of vaccination starting 2020-12-27 in Sweden. 

The data used in our analysis is publicly available from the website of the Public Health Agency of Sweden~\cite{FHM}, where new reports have been published daily from Tuesday to Friday (excluding public holidays). The aggregated daily counts are updated retrospectively at each reporting date. As the case fatalities are associated with a reporting delay, this implies that the published time series of reported COVID-19 fatalities will always show a declining trend (see Fig~\ref{fig:rep_vs_unrep} for an illustrative example). The reporting delay can not be observed in a single published report but can be obtained by comparing the aggregated numbers of fatalities of each date from previously published reports.

\subsection*{Nowcasting}
\begin{figure}[!h]
\centering
\includegraphics[width=0.8\textwidth]{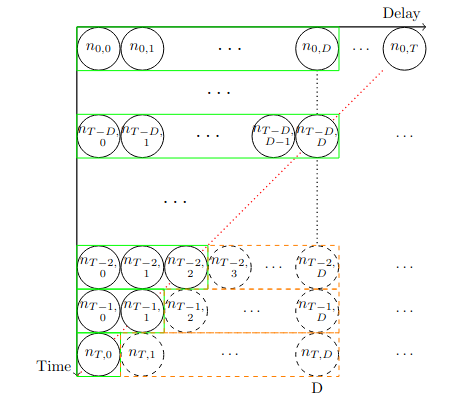}
\caption{{\bf Reporting triangle for day \textit{T}.}
 Green boxes (solid line) where $t \le T - D$ are the reported number of fatalities on day $T$ (today), considering a maximum delay of $D$ days. The red boxes (dashed line), corresponding to $t > T - D$, are the occurred-but- not-yet-reported number of events of day $t+D$. }
\label{fig:rep_tri}
\end{figure}

The notation and methodological details of our approach follows closely the notation introduced in G{\"u}nther et al.~\cite{gunther_2020}. Let $n_{t,d}$, be the number of fatalities occurring on day $t=0,...,T$ and reported with a delay of $d=0,1,2,...$ days, such that the reporting occurs on day $t+d$. The goal of Nowcasting is to infer the total number of fatalities $N_t$ of day $t$ based on the information available on the current day $T \ge t$. The sum $N_t$ can be written as
\begin{eqnarray*}
    N_t =\sum_{d=0}^{\infty}n_{t,d}= \sum_{d=0}^{T-t}n_{t,d} + \sum_{d=T-t+1}^{\infty}n_{t,d},
\end{eqnarray*}
where the first sum is observed and the second sum is yet unknown. This can be illustrated by the so called reporting triangle (Fig~\ref{fig:rep_tri}). Where the upper left triangle are the number of reported fatalities and the lower right triangle is the number of occurred- but-not-yet-reported events with a maximum delay of \textit{D} days. The upper triangle carries the information about the reporting delay from the past and the lower triangle is what is estimated with the Nowcasting model. 

We let $\lambda_t$ denote the expected value of $N_t$, and $p_{t,d}$ denote the conditional probability of a fatality occurring on day $t$ being reported with a delay of $d$ days. Then, the number of events occurring on day $t$ with a delay of $d$ days is assumed to be negative binomial distributed 
\begin{eqnarray*}
    n_{t,d}|\lambda_t, p_{t,d} \sim  \text{NB}(\lambda_t \cdot p_{t,d}, \phi),
\end{eqnarray*}
with mean $\lambda_t \cdot p_{t,d}$ and overdispersion parameter $\phi$. Hence, the Nowcasting task can be seen as having two parts; (1) determine the expected value of the total number of fatalities and (2) determine the reporting delay distribution to subsequently predict the $n_{t,d}$'s and finally compute the $N_t$'s.

\subsection*{Flexible Bayesian Nowcasting}
As described in the previous section, the nowcasting problem can be seen as a problem of the joint estimation of two models: (1) a model for the expected number of deaths over time, and (2) a model for the reporting delay distribution, which can also vary over time. Therefore, we let our model constitute of two distinct elements;( 1) the underlying epidemic curve determining the expected number of fatalities $\lambda_t$ and (2) the reporting delay distribution determining $p_{t,d}$. We will in the following describe the structure of each.

\subsubsection*{Component 1: The expected number of fatalities}\label{sec:mod_comp1}
Let $\lambda_t=\mathbbm{E}[N_t]$ denote the expected total number of fatalities occurring on day $t$. 
We specify a baseline model for $\lambda_t$ as
\begin{eqnarray}\label{eq:model1}
    \log(\lambda_t)|\lambda_{t-1} \sim N(\log(\lambda_{t-1}), \sigma^2),
\end{eqnarray}
where $t=0,...,T$ and $d=0,...,D$. Time $t=0$ is assumed to be the start of the chosen observation period, e.g. the start of the pandemic. This approach to model $\lambda_t$ as a Random Walk on the log scale is proposed by McGough et al.\cite{mcgough_etal2020} and G{\"u}nther et al.\cite{gunther_2020}. Here, we will refer to it as model R.

An alternative to model R in Eq~\eqref{eq:model1} is to assume that we can predict the total number of fatalities with additional data streams associated with the event of interest. The additional data streams are assumed to be ahead in time compared to the time series of interest, e.g.\ due to the tracked event of the stream being at an earlier stage in a typical COVID-19 disease progression or because of a smaller reporting delay; e.g.\ the number of reported cases and hospitalizations, etc. Hence, we can use the additional data stream as a leading indicator in the Nowcasting model. One approach is to consider the number of fatalities as some time-varying fraction of the numbers in those additional data streams. We denote the $i$'th of the $k$ leading indicators at time $t$ as $m_{i,t}$ and specify an regression type model for $\lambda_t$ as follows:
\begin{equation}\label{eq:model3}
   \log(\lambda_t)|m_{1,t},...{m_{n,t}}  \sim N\left(\beta_0 + \sum_{i=1}^k \beta_{i}  m_{i,t}, \sigma^2\right),
\end{equation}
where the $\beta_0$ is an intercept and $\beta_i$ denotes the additive effect of the $i$'th stream on the log of the mean of $\lambda$. With this model specification, we assume a strong association between the case fatalities and the $k$ data streams suitably measured some days earlier. We will refer to this model as L($m_i$).

Furthermore, we propose another approach combining the random walk component of the model in Eq~\eqref{eq:model1} and the additional data streams of Eq~\eqref{eq:model3}. Here, we let the leading indicators be the relative change in e.g. case reports or hospitalizations. In other words, we assume that if there is an increase in the leading indicator, we also expect an increase in the number of fatalities. An increase in case reports is not expected to give an instant increase in the number of deaths but rather with some time delay, so as for the model in Eq~\eqref{eq:model3}, the leading indicators need to be specified with a suitable time delay. We specify this alternative model for $\lambda_t$ as
\begin{equation}\label{eq:model2}
    \log(\lambda_t)|\lambda_{t-1}, m_{1,t},...,m_{n,t} \sim N\left(\log(\lambda_{t-1})+\sum_{i=1}^k\beta_i m_{i,t}, \sigma^2\right),
\end{equation}
where the $\beta_i$'s are again considered as regression coefficients for the leading indicator $m_i$. This approach combines an established method \cite{gunther_2020} with additional information that is informative of the events of interest. We note that when the $\beta$-coefficients of this model are zero, this model becomes identical to the model specified in Eq~\eqref{eq:model1}. This model will be referred to as RL($m_i$).

\subsubsection*{Component 2: The reporting delay distribution}

The model for the reporting delay distribution at day $t$ is specifying the probability of a reporting delay of $d$ days for a fatality occurring on day $t$. We denote this conditional probability 
\begin{eqnarray*}\label{eq:p_td}
p_{t,d}= P(\text{delay}=d|\text{fatality day} = t).
\end{eqnarray*}
Similarly to G{\"u}nther et al.~\cite{gunther_2020}, we model the delay distribution as a discrete time hazard model $h_{t,d}=P(\text{delay}=d|\text{delay}\ge d, W_{t,d})$ as
\begin{equation}\label{eq:h_t}
    \text{logit}(h_{t,d})=\gamma_d+W'_{t,d}\eta,
\end{equation}
 where $d=0,...,D-1, h_{t,D}=1$, $\gamma_d$ is a constant, $W_{t,d}$ being a vector of time- and delay-specific covariates and $\eta$ the covariate effects. It can be shown how the reporting probabilities are derived from Eq~\eqref{eq:h_t}~\cite{gunther_2020}.
 We are using linear effects of the time on the logit-scale with break-points every two weeks before the current day to allow for changing dynamics in the reporting delay distribution over time. We also use a categorical weekday effect to account for the weekly structure of the reporting.
 \subsection*{Inference and implementation}

 Inference for the hierarchical Bayesian nowcasting model is done by Markov Chain Monte Carlo using R-Stan \cite{rstan_2020} extending the work of G{\"u}nther et al.~\cite{gunther_2020}.  In order to ensure reproducibility and transparency, the R-Code~\cite{r_2021} and data used for the analysis is available from \url{https://github.com/fannybergstrom/nowcasting\_covid19}.

\section*{Results}

\subsection*{Application to fatalities}\label{application} 

We apply the nowcasting methods to reported COVID-19 fatalities in Sweden and let the number of reported cases and COVID-19 associated ICU admissions act as two leading indicators. In Sweden, the reporting of ICU admissions is also associated with a reporting delay but considerably shorter than the fatalities. We use model R as a benchmark model and compare it the two alternative models using leading indicators; model L where we let the leading indicator be the number of COVID-19-related ICU admissions, and model RL including both the random walk component and leading indicator here being the relative weekly change in ICU admissions. We denote the leading indicator models as L(ICU) and RL(ICU). For the leading indicator time series, we use a seven day centered rolling average to avoid the weekday effect of the reporting. The pre-specified lag between the fatalities and leading indicators is determined by fitting a linear time series model given the two model specifications of models L and RL, and choosing the lag providing the best fit. The period chosen for the time series model is 2020-04-01--2020-10-19 to use the information available only prior to the evaluation period. We use 18 days lag for the reported cases and 14 days for the ICU admissions. The reporting probability is set to be zero on non-reporting days (Saturday--Monday and public holidays). Furthermore, for practical and robustness reasons, a maximum delay of $D=35$ days is considered. For the fatalities reported with a longer than the maximum, we set their delay to the upper limit of 35 days. There were 116 case fatalities reported with a delay longer than 35 days during the evaluation period. 

\subsection*{Retrospective nowcasting evaluation}
We use a retrospective evaluation in order to assess the performance of the Nowcasting models. The model-based predictions are compared to the (now assumed to be known) final number of COVID-19-related reported fatalities in Sweden. The samples from the posterior predictive distribution for the total number of reported COVID-19 fatalities $\hat N_t$ are extracted for each reporting date of the evaluation period. 
As in G{\"u}nther~\cite{gunther_2020}, we use the following four metrics to quantify the model performance; 1) continuous rank probability score (CRPS), 2) log scoring rule (logS), 3) root mean squared error (RMSE), and 4) the prediction interval (PI) coverage being the proportion of times the true number of fatalities is contained within the equitailed PI. The RMSE is calculated with a point estimate being the median of the posterior predictive samples of $\hat N_t$, while the scoring rules CRPS and logS assess the quality of the probabilistic forecast by considering the full posterior distribution of $\hat N_t$~\cite{Gneiting2007StrictlyPS}. For the scoring rules, a low score indicates a better performance.
 
\begin{figure}[H]
\includegraphics[width=1\textwidth]{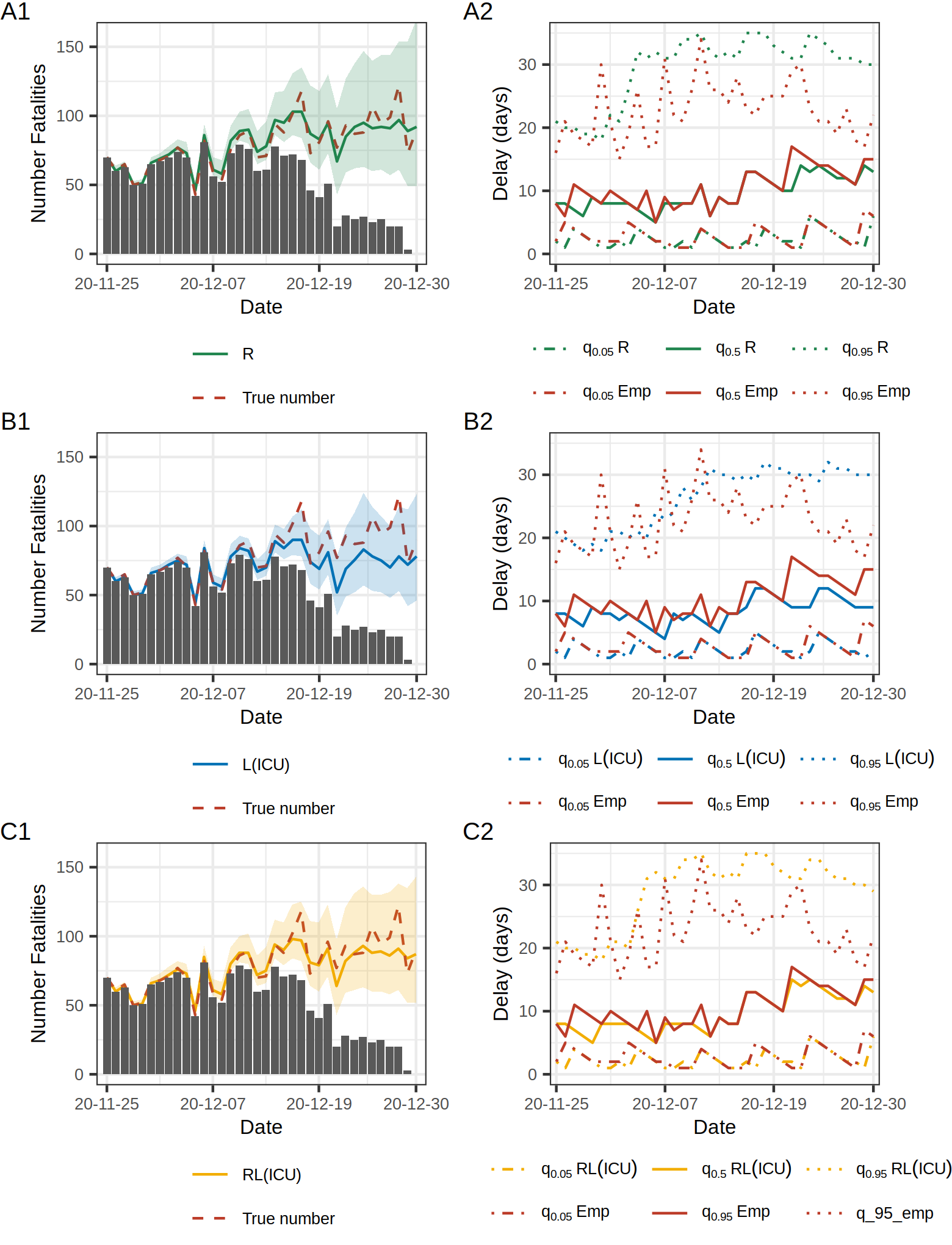}
\vspace{1mm}
\caption{{\bf Nowcasts for a specific reporting date 
}
Left column shows the nowcasts of 2020-12-30, where the solid lines are the median of the posterior predictive distribution of $\hat N$ and the shaded area depict the 95\% PI. The black bars are what is yet reported and the red line is the true number, only known retrospectively. Right column shows quantiles of the estimated and empirical reporting delay distribution. The solid lines the median reporting the delay in days (for each date) and the lower and upper bounds are the 5\% and 95\% quantiles. At the 5\% quantile, 5\% of the total number of fatalities occuring on that date are estimated to be reported within the number days delay, ect. The empirical quantiles are obtained with data available in hindsight.}
\label{fig:snapshot_res}
\end{figure}

Nowcasts and the estimated reporting delay for a specific reporting date \textit{T}=2020-12-31, is shown in Fig~\ref{fig:snapshot_res}. In the left column, the black bars are the number of fatalities reported until day $T$ and the red dashed line is the true number, only known in retrospect. The solid lines are the median of the posterior predictive distribution of $\hat N_t$ and the shaded areas indicate the equitailed point-wise 95\% Bayesian prediction interval, estimated with information available at the reporting date. The right column shows the daily empirical and estimated number of days of reporting. The solid lines are the estimated and empirical median days of reporting delay and the shaded area is between the 5\% and 95\% quantile of the reporting delay. The lower bound indicate the number of days until 5\% of the total number of fatalities will be reported and the upper bound is within how many days 95\% will be reported. The empirical median and the respective quantiles are calculated with data available in hindsight and the estimated quantities are obtained with the information available at the reporting date.

We observe an underestimation of the reporting delay for the L(ICU) model for the last days in the observation window (2020-12-25--2020-12-30) resulting in an underestimation of the daily number of fatalities (Fig~\ref{fig:snapshot_res}B). We can also note that the PI is more narrow for L(ICU) than for the other two models and that the true number is not always contained in the PI. Model R and RL(ICU) (Fig~\ref{fig:snapshot_res}A \& C) provide similar results with less underestimation of the reporting delay resulting in a point estimate of the median of the predictive distribution lying closer to the true number compared to model L(ICU). A difference between the performance between R and RL(ICU) is that RL(ICU) provides less wide PI than R. For R and RL(ICU), the true number of daily fatalities is contained in the PI for all days \textit{T-t}, $t=0,\dots,35$.
From the right column of the figure, it can be observed that the 5\% quantile of the estimated number of days of reporting delay for all three models are similar to the empirical 5\% quantile. Also, the median of the estimated number of days delay follows the corresponding empirical quantity reasonably well. Contrary, the 95\% estimated quantiles are farther from the empirical. This indicate that all three models capture the short-term trends such as the weekly reporting patterns well but do not fully capture the changing dynamics of the long reporting delays, i.e.~the high spikes in the early observation window and the rapid decrease in the final week. An alternative visualization of the empirical and estimated reporting delay distribution for the three models provided by the cumulative reporting probability is found in \nameref{S1_Appendix}~Sec 1. 

\begin{figure}[H]
\includegraphics[width=1\textwidth]{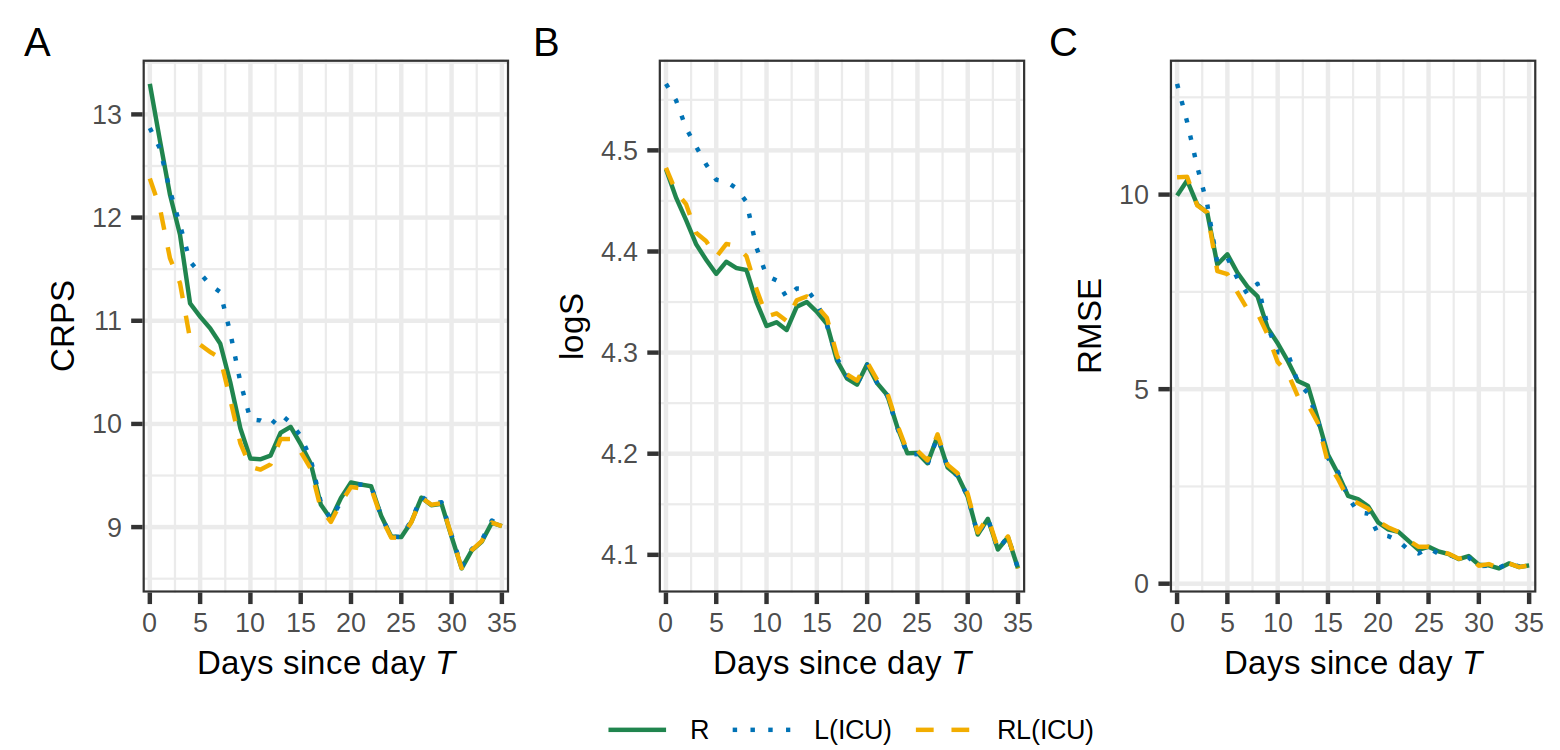}
\vspace{1mm}
\caption{{\bf Mean scores by the number of days since the day of reporting \textit{T}.} The results are averaged over all reporting dates $T$ in the evaluation period from 2020-10-20 to 2021-05-21.
}
\label{fig:S2}
\end{figure}

Seen in Fig~\ref{fig:snapshot_res}, the PI is increasing as the final date \textit{T} of the observation window is approaching. As the number of days \textit{t} since day \textit{T} decrease, the uncertainty for the nowcast of day \textit{T-t} increase as the fraction of the reported fatalities will be decreasing. The average score as a function of \textit{T-t} is shown in Fig~\ref{fig:S2}. For all models and scores, the score is generally a decreasing function of the number of days since day \textit{T}. Hence, the farther from \say{now}, the closer are the nowscast estimates of the daily number of fatalities to the true number. The difference in performance for the three models is observable for the two weeks prior to day \textit{T}. Here, we see that model RL(ICU) has a lower CRPS and RMSE score (Fig~\ref{fig:S2}A \& C) and that model R has the lowest logS (Fig~\ref{fig:S2}B). Model L(ICU) has the overall highest values of the scores, hence it has the worst performance of the three models. 

The mean overall score and the coverage frequency of the 75\%, 90\%, and 95\% prediction interval of the three models for the nowcasts performed in the evaluation period is found in  Table~\ref{table1}. For each reporting day $T$, we consider the average score of the last seven days; $T-6,\dots,T-0$.
Based on the CRPS and RMSE, model RL(ICU) has the best performance, with a decrease of 4.2\% and 1.0\% respectively compared to model R. Model R has the lowest logS score but only with a slight advantage compared to RL(ICU) (0.02\% improvement). Model L(ICU) has the worst performance for all three scores. The coverage of the prediction intervals for models R and RL(ICU) is of satisfactory levels. In contrast, the L(ICU) model has low coverage, indicating that model L(ICU) is less trustworthy. 
\begin{table}[!ht]
\centering
\caption{
{\bf Results of the retrospective evaluation of different nowcasting models on COVID-19 related fatalities in Sweden.}}
\begin{tabular}{|l+l|l|l|l|}
\hline
 \textbf{Score} &  \textbf{R} &  \textbf{L(ICU)} &  \textbf{RL(ICU)} \\
\thickhline
 CRPS &  11.89   & 12.01 & \textbf{11.39}\\
   \hline
 logS & \textbf{4.42}   & 4.51 & 4.43 \\
 \hline
    RMSE & 9.18 &  9.95 & \textbf{9.09} \\  
    \thickhline
     Cov. 75\% PI &  76.07\%  & 58.97\% & 76.07\%\\
   \hline
 Cov. 90\% PI &  95.72\%  & 76.07\% & 94.87\%\\
  \hline
 Cov. 95\% PI & 98.29\% & 84.61\%  & 99.14\%  
 \\ \thickhline
\end{tabular}
\begin{flushleft} CRPS is the continuous ranked probability score, logS is the log score, and RMSE denotes the root mean squared error of the posterior median. Additionally, we provide coverage frequencies of 75\%, 90\% and 95\% credibility intervals in the estimation of the daily number of case fatalities. The scores are averaged over nowcasts for day $T-6,...,T-0$, with $T$ being all reporting dates in the evaluation period.
\end{flushleft}
\label{table1}
\end{table}

\begin{figure}[!h]
\includegraphics{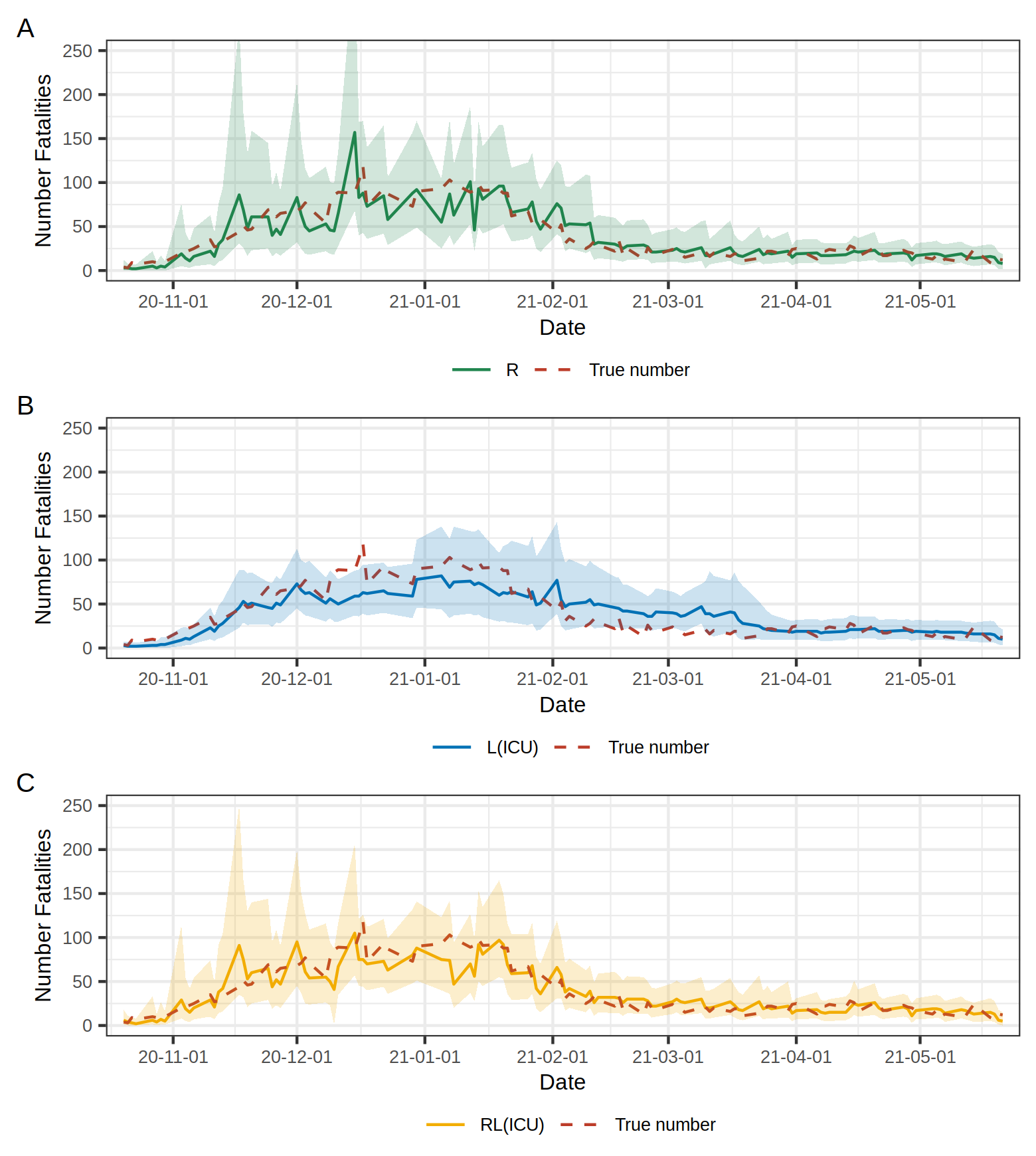}
\caption{{\bf Estimated and true number of fatalities with COVID-19 in Sweden.} 
The estimated number of fatalities are the nowcasts of day $T$ being each reporting date in the evaluation period from 2020-10-20 to 2021-05-21. The solid lines are the median of the posterior predictive distribution of the number of daily fatalities $\hat N_T$ and the shaded area depict the point-wise 95\% PI. The red line is the retrospective true number.}
\label{fig6}
\end{figure}

Fig~\ref{fig6} shows the retrospective true number of daily fatalities and the median of the predictive distribution of $\hat N$ and a 95\% PI of day $T$ for the three models evaluated on each reporting day in the evaluation period. In Fig~\ref{fig:snapshot_res}, this corresponds to the nowcast estimates of the final date \textit{T}=2020-12-30. We observe a similar performance over time for models R and RL(ICU) (Fig~\ref{fig6}A \& C) and the more significant deviations from the true number appear mainly on the same reporting dates for the two models. In early Jan 2021, RL(ICU) underestimates the number of daily fatalities, likely due to the rapid decrease in ICU admissions due to the introduction of vaccines at the end of Dec 2020, while the case fatalities were also on a downwards trend but not as steep. Model RL(ICU) stabilizes after approximately two weeks (same as the length of the linear change points) in mid Jan 2021 as the model adapts to the new association between ICU admissions and case fatalities. Model L(ICU) (Fig~\ref{fig6}B) does not have the high peaks in the posterior predictive distribution of $\hat N$ as the other two models. However, the deviation of the posterior median compared to the true number is visibly larger. Starting from Dec 2020, we observe an underestimation of the number of fatalities, and from Feb 2021, an overestimation for the following two months. 
From Apr 2021 until the end of the evaluation period, the three models have a visibly similar performance with a posterior mean close to the true number of daily fatalities and a narrow PI containing the true number.

The performance of the alternative models with leading indicators compared to model R can be explained by the estimated association between the fatalities and the leading indicators. The changing dynamics of the association over time are captured by the estimated time-varying $\beta$-coefficients of the respective models. Details of the estimated $\beta$-coefficients for models R(ICU) and RL(ICU) over the evaluation period are reported in~\nameref{S1_Appendix} Sec 2.
 
\begin{figure}[!h]
\includegraphics{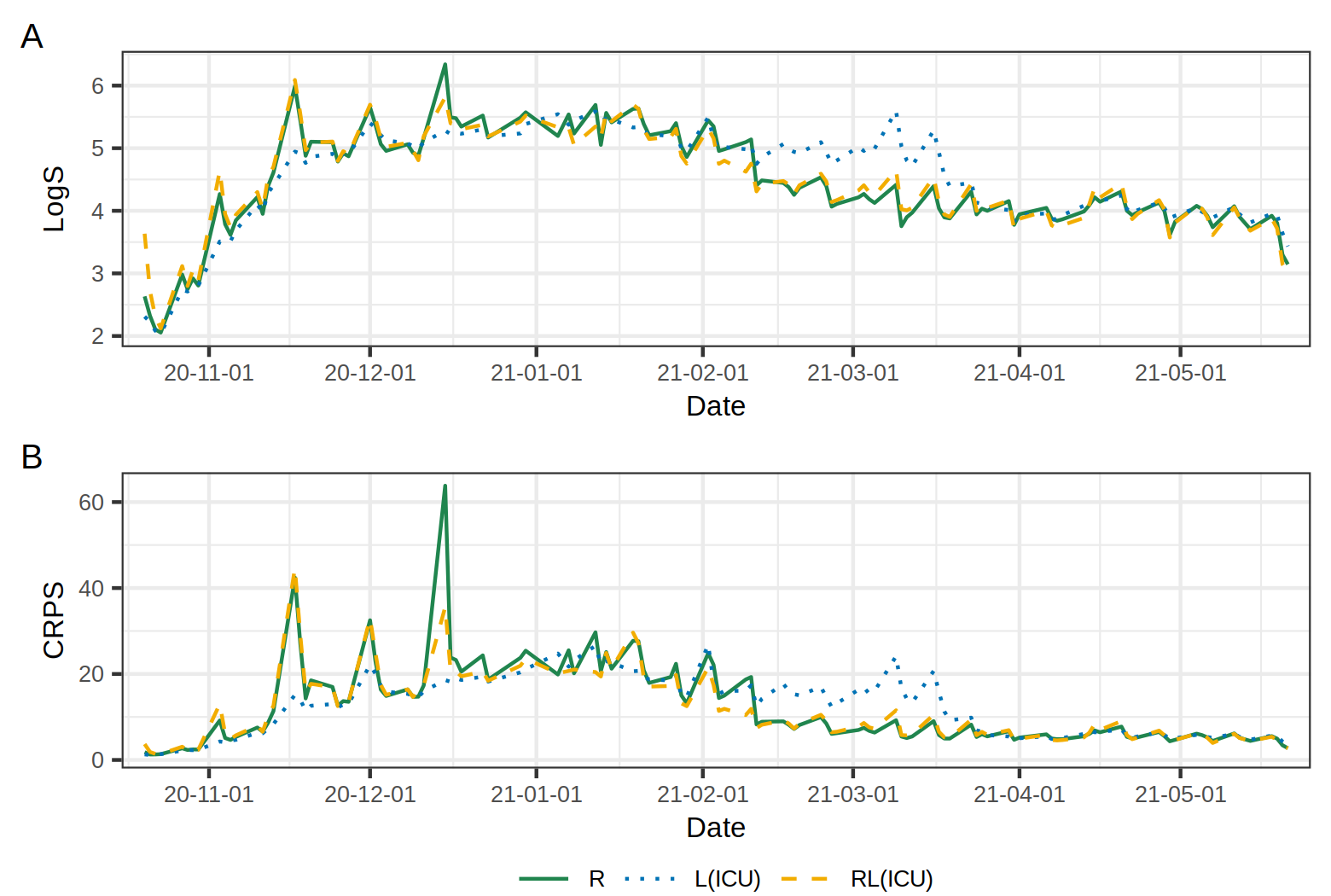}
\vspace{2mm}
\caption{{\bf Scoring rules.} Average CRPS and logS of the last 7 days; $T-6,\dots,T-0$ for each reporting day $T$, in the evaluation period.}
\label{fig:score7}
\end{figure}

The scores of the three models evaluated at the 117 reporting dates in the evaluation period by the CRPS and LogS is shown in Fig~\ref{fig:score7}. For each reporting day $T$, we consider the average score of the last seven days; $T-6,\dots,T-0$. 
For the three models, the scores are generally higher when the number of case fatalities is high. Overall, the performance of model R and RL(ICU) is similar, as could also be observed in Fig~\ref{fig6}.
From the beginning of the evaluation period until the end of 2020, model L(ICU) has an overall lower score and a more stable performance with less high spikes in the score compared to model R and RL(ICU). During Jan 2021, the performance is similar for the three models, but from Feb to Apr 2021 model L(ICU) performs significantly worse than the other models. The remaining scoring rule, the RMSE, entail similar results (\nameref{S:rmse}). After Apr 2021, the number of daily fatalities has stabilized to a low number and the score for three models becomes similar until the end of the evaluation period. 

In conclusion, we find that model R and model RL(ICU) performs well over the evaluation period and has a satisfactory level of PI coverage. Furthermore, model RL(ICU) provided the best performance of the three models, indicating that there is a gain (4.2\% decrease in CRPS compared to model R) of including leading indicators. Using reported cases or the combination of reported cases and ICU admissions as leading indicators does not improve performance. The results of using these leading indicators are found in~\nameref{S1_Appendix} Sec 3.

\section*{Discussion}

In the presented work, we provide an improved method for real-time estimates of infectious disease surveillance data suffering a reporting delay. The proposed method can be applied to any disease for which the data can be put the form of the reporting triangle given in Fig~\ref{fig:rep_tri}. 
We apply the method to COVID-19-related fatalities in Sweden. Even though fatalities are a lagging indicator to obtain situational awareness about the pandemic and is not without difficulties itself, it is often used as a more robust indicator to assess the burden of disease because it might be less influenced by the current testing strategy. Hence, monitoring the time series of reported deaths has been of importance in the still on-going COVID-19 pandemic. 

We show that using leading indicators, such as the COVID-19-associated ICU admissions, can help improve the nowcasting performance of case fatalities compared to existing methods. 
Beyond using reported cases and ICU admissions as leading indicators for the case fatalities, other possible leading indicators are e.g.~vaccination, hospitalizations, and virus particles in wastewater~\cite{kreier_2021}, or using age-stratified reported cases. However, nowcasting with leading indicators should be made with caution and be reevaluated as the dynamics between the leading indicator and the event of interest change, which may not be a trivial task during an ongoing pandemic.
Furthermore, by re-estimating the association coefficients of the leading indicator at each reporting date, our method captures the changing association between ICU admissions and case fatalities over time. However, we use a pre-specified time lag unknown at the start of the pandemic and might also change throughout the pandemic. A possible extension of our work would thus be to estimate this time lag as a part of the model fitting.

The proposed method is flexible in terms of its application and thus can be a helpful tool for future pandemic stress situations. We support this by providing open-source software for the real-time analysis of surveillance data. Weekly updated nowcast estimates of COVID-19 fatalities and ICU admissions in Sweden using our proposed method, model RL, are found at 
\begin{center}
    \url{https://staff.math.su.se/fanny.bergstrom/covid19-nowcasting} 
\end{center}
These graphs help provide the desired situational awareness and are to be interpreted as new variants emerge.

\section*{Supporting information}

\paragraph*{S1 Fig.}
\label{S:rmse}
{\bf RMSE.} Average RMSE of the last 7 days; $T-6,\dots,T-0$ for each reporting day $T$, in the evaluation period.


\paragraph*{S1 Appendix.}
\label{S1_Appendix}
{\bf Complimentary material and results.} Sec 1 contains information about the cumulative reporting probability, providing a complimentary picture of the estimated reporting delay. Sec 2 presents detailed results of the estimated regression coefficients of model L(ICU) and RL(ICU) over the evaluation period. Finally, Sec 3 covers results of including reported cases and the combination of reported cases and ICU admissions as leading indicators. 

\section*{Acknowledgments}
This work is partly funded by the Nordic Research Agency
. The computations and data handling was enabled by resources provided by the Swedish National Infrastructure for Computing (SNIC) at HPC2N partially funded by the Swedish Research Council. 
We also thank Markus Lindroos for discussions and his contribution in coding of the reporting delay distribution.

\nolinenumbers

%
%
%

\bibliography{nowcasting}








\section*{Supplementary material}

# SI Appendix
# Flexible Bayesian Nowcasting with leading indicators applied to COVID-19 fatalities in Sweden

July 12, 2022

## 1 Cumulative reporting probability

The cumulative reporting probability is the proportion of reported case fatalities until a certain amount of time. E.g. if the cumulative probability for 10 days is 40%, it means that 40% of the cases will be reported within 10 days. In Fig S1, the cumulative reporting probability is shown for the nowcasts estimates for the three models R, L(ICU), and RL(ICU) of 2020-12-30, together with the empirical fraction of reported cases (known in hindsight).

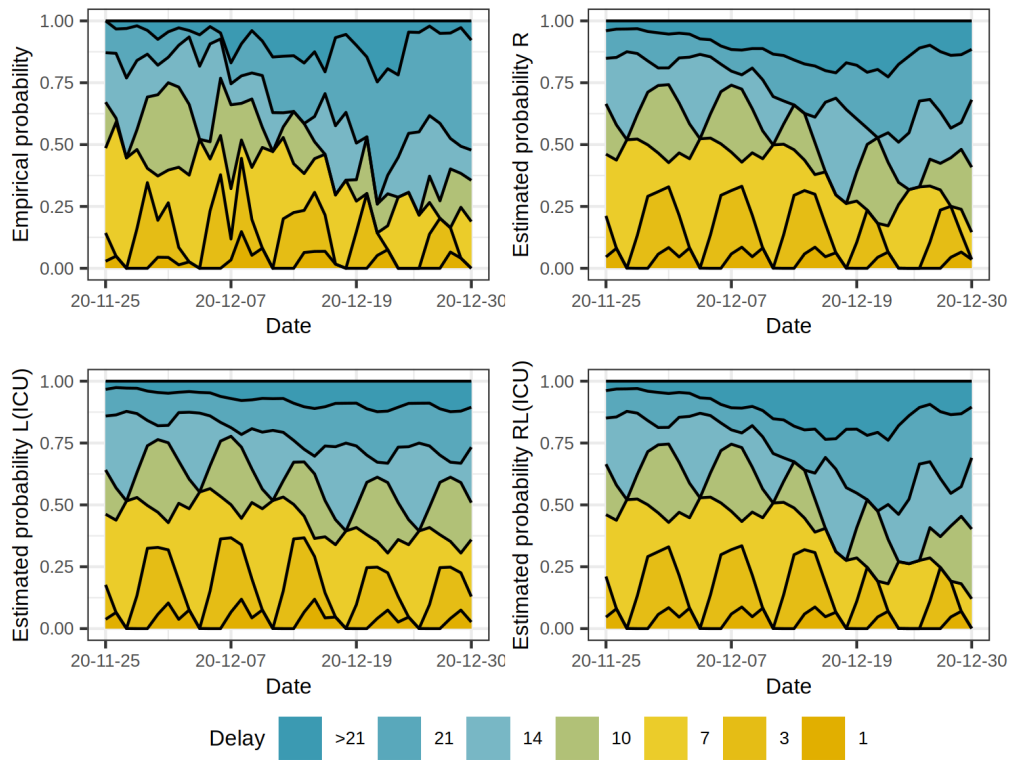

Figure S1: Cumulative reporting probability of the nowcast model of 2020-12-30 for the three models; R, L(ICU), and RL(ICU), as well as the empirical reporting probability (known in hindsight). For a specific date, the cumulative reporting probability is the fraction of case fatalities occurring on that date reported within a given number of days.

1

The empirical fraction of the reported fatalities shows a clear weekly pattern but with high irregularities in the reporting from week to week. One can also note the increase in the reporting delay over time. The three nowcasting models manage to capture the overall weekly pattern as well as the increase in the reporting delay. The cumulative reporting probability is complimentary to Fig 4 in the sense that it also illustrated the estimated reporting delay distribution for the period 2020-11-25–2020-12-30, estimated by the nowcast models with the information available as of 2020-12-30. One can observe that model L(ICU) underestimate the reporting delay for the last week of the observation window, leading to an underestimation of the case fatalities for this period. In contrast to model R and RL(ICU) that manage to capture the increase in reporting delay to a higher extent. No case fatalities were reported within the same day (with a delay of zero days) for this period.

## 2 Regression coefficients

The estimated regression coefficients for model L(ICU) and RL(ICU) for each reporting date in the evaluation period is shown in Fig S2 and S3.

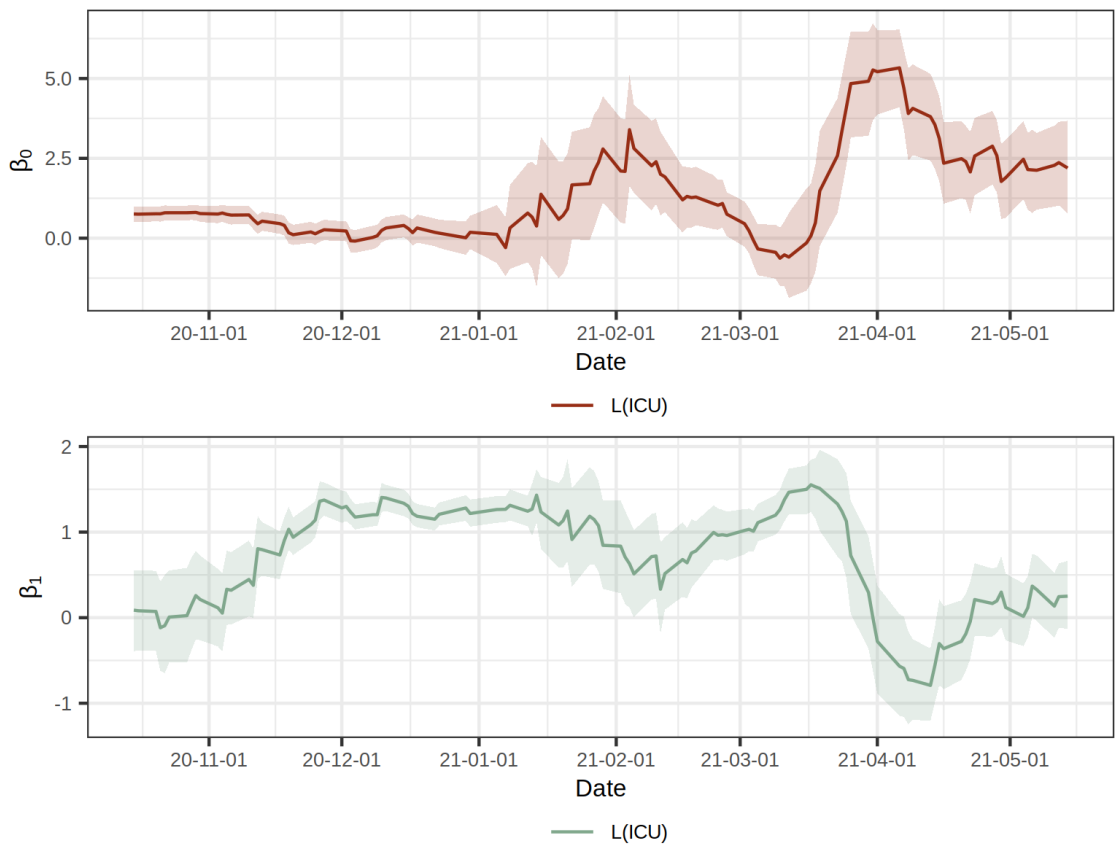

Figure S2: Estimated regression coefficients for L(ICU). The solid lines are the median of the posterior predictive distribution of $\beta_0$ and $\beta_1$ and the shaded areas indicate the equitailed pointwise 95% Bayesian prediction interval.

In the beginning of the evaluation period, the intercept $\beta_0$ of model L(ICU) is stable at a low level and the regression coefficient $\beta_1$ is increasing as both ICU admissions and case fatalities are increasing. In Jan 2021 when the vaccination is introduced, the association becomes less strong and the $\beta_1$ coefficient is decreasing and instead we see an increase of the intercept. When the ICU

admissions start to rise again due to the delta wave, the association also rises, but this time on false premises since the association now is less strong, resulting in the bad performance of this model in the period of 2021-02-15–2021-03-15. Eventually, the new association are captured by the model and the final two weeks of the evaluation period the model has a stable intercept $\beta_0$ and a low estimate of $\beta_1$.

The estimated regression coefficient $\beta_1$ of model RL(ICU) (Fig S3), where the leading indicator is the relative weekly change in ICU admissions, is more constant until the introduction of vaccines. After that both ICU admission and fatalities and drop, and the estimated association stays positive. During the delta wave, the case fatalities continue to decrease while the ICU admissions start to rise, hence the estimated negative association during April 2021. The final two weeks of the observation period, the median of the posterior distribution of $\beta_1$ is centered around zero.

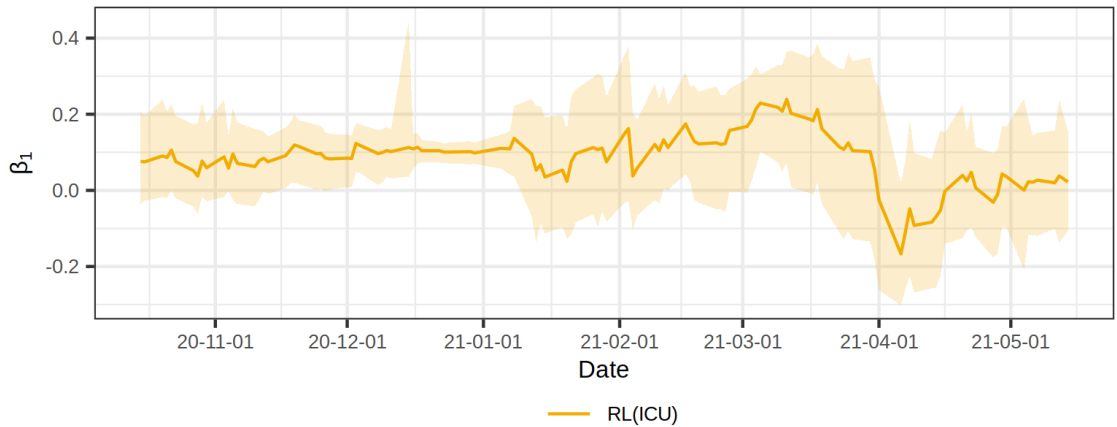

Figure S3: Estimated regression coefficient for RL(ICU). The solid lines are the median of the posterior predictive distribution of $\beta_1$ and the shaded area indicate the equitailed point-wise 95% Bayesian prediction interval.

## 3 Multiple data streams

In addition to using ICU admission as a single leading indicator, we also used reported cases and the combination of reported cases and ICU admissions as leading indicators. In Fig S4, the result of the model RL(Cases) and RL(Cases, ICU) is shown.

Compared to model RL(ICU), the performance of models RL(Cases) and RL(Cases, ICU) is worse. The association between the reported cases and the case fatalities is lower than for the ICU admissions, which explains that there is no gain in including the reported cases as a leading indicator. The time series of reported cases heavily depends on the reporting capacity and strategy, as well as the propensity of the general public to get tested, which all has been changing rapidly under the course of the pandemic. The association between the ICU admissions and case fatalities has been more stable throughout the pandemic with the exception of the introduction of the vaccines in Jan 2021.

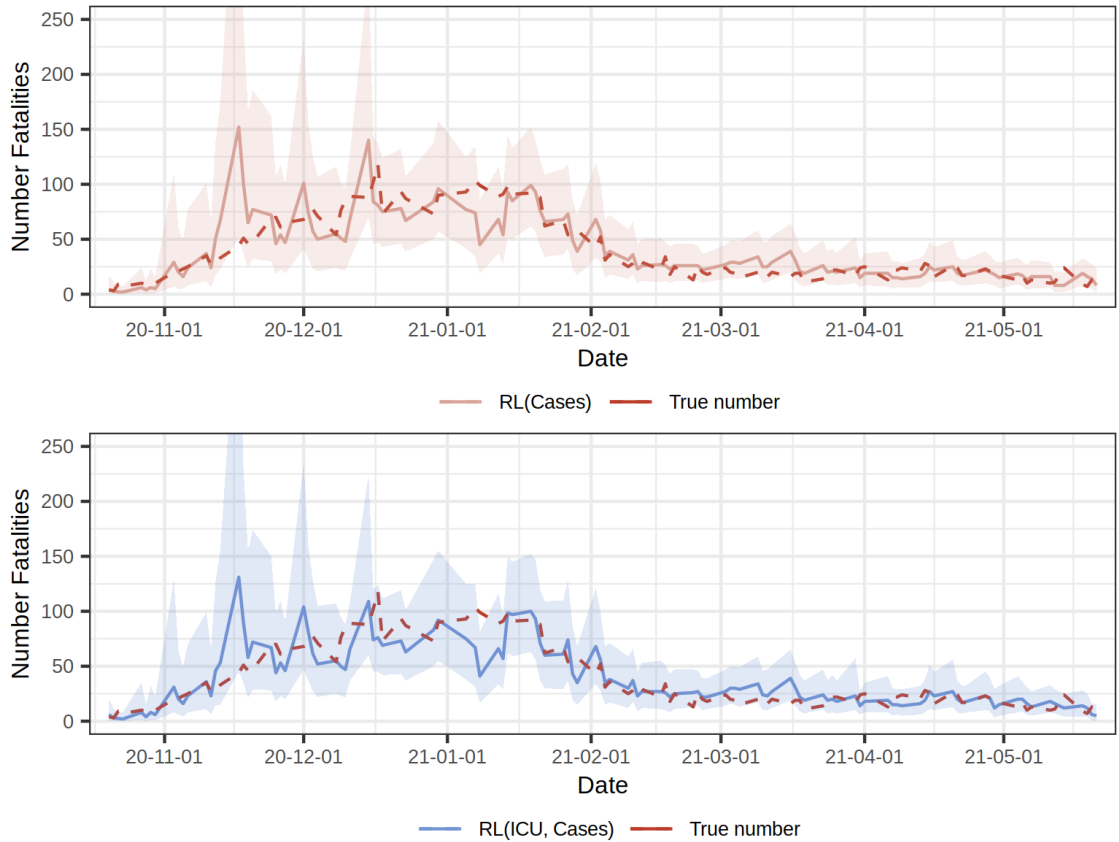

Figure S4: Median of the posterior predictive distribution of $\hat{N}$ (solid line) and a 95% PI, and the actual number number of fatalities (known in hindsight).

4



\end{document}